\documentclass{article}

\usepackage{booktabs}
\usepackage{graphics}
\usepackage{graphicx}
\usepackage{listings}
\usepackage{nameref}
\usepackage{algorithm}
\usepackage{algpseudocode}
\algrenewcommand\algorithmicrequire{\textbf{Input:}}
\algrenewcommand\algorithmicensure{\textbf{Output:}}
\usepackage{bm}
\usepackage{amsmath}
\usepackage{amssymb}
\usepackage{caption} 
\usepackage{multirow}

\newcommand{\name}{PhaMer}
\newcommand{\Cherry}{PhaMer}
\usepackage{arxiv}

\usepackage[utf8]{inputenc} 
\usepackage[T1]{fontenc}    
\usepackage{hyperref}       
\usepackage{url}            
\usepackage{booktabs}       
\usepackage{amsfonts}       
\usepackage{nicefrac}       
\usepackage{microtype}      
\usepackage{lipsum}
\usepackage{graphicx}
\graphicspath{ {./images/} }

\title{Accurate identification of bacteriophages from metagenomic data using Transformer}

\author{
 Jiayu Shang \\
  Dept. of Electrical Engineering\\
  City University of Hong Kong\\
  Kowloon, Hong Kong SAR, China\\
  \texttt{jyshang2-c@my.cityu.edu.hk} \\
  \And
 Xubo Tang \\
  Dept. of Electrical Engineering\\
  City University of Hong Kong\\
  Kowloon, Hong Kong SAR, China\\
  \texttt{xubotang2-c@my.cityu.edu.hk} \\
  \And
  Ruocheng Guo\\
  School of Data Science\\
  City University of Hong Kong\\
  Kowloon, Hong Kong SAR, China\\
  \texttt{ruocheng.guo@cityu.edu.hk} \\
  \And
 Yanni Sun \\
  Dept. of Electrical Engineering\\
  City University of Hong Kong\\
  Kowloon, Hong Kong SAR, China\\
  \texttt{yannisun@cityu.edu.hk} \\
}

\begin{document}

\maketitle
\begin{abstract}
\textbf{Motivation:} \textbf{Motivation:} Bacteriophages are viruses infecting bacteria. Being key players in microbial communities, they can regulate the composition/function of microbiome by infecting their bacterial hosts and mediating gene transfer. Recently, metagenomic sequencing, which can sequence all genetic materials from various microbiome, has become a popular means for new phage discovery. However, accurate and comprehensive detection of phages from the metagenomic data remains difficult. High diversity/abundance, and limited reference genomes pose major challenges for recruiting phage fragments from metagenomic data. Existing alignment-based or learning-based models have either low recall or precision on metagenomic data. \\
\textbf{Results:} In this work, we adopt the state-of-the-art language model, Transformer, to conduct contextual embedding for phage contigs. By constructing a protein-cluster vocabulary, we can feed both the protein composition and the proteins' positions from each contig into the Transformer. The Transformer can learn the protein organization and associations using the self-attention mechanism and predicts the label for test contigs. We rigorously tested our developed tool named \Cherry\ on multiple datasets with increasing difficulty, including quality RefSeq genomes, short contigs, simulated metagenomic data, mock metagenomic data, and the public IMG/VR dataset. All the experimental results show that PhaMer outperforms the state-of-the-art tools. In the real metagenomic data experiment, PhaMer improves the F1-score of phage detection by 27\%. \\
\textbf{Availability:} The source code and supplementary file of \Cherry\ is available via: \href{https://github.com/KennthShang/\Cherry}{https://github.com/KennthShang/\Cherry}\\
\textbf{Contact:} \href{yannisun@cityu.edu.hk}{yannisun@cityu.edu.hk}\\
\end{abstract}


\newpage

\section{Introduction}
\label{sec:intro}
Bacteriophages (phages for short) are viruses infecting bacteria. They are highly ubiquitous and are widely regarded as the most abundant organisms on Earth \cite{mcgrath2007bacteriophage}. There is accumulating evidence showing phages' significant impacts on various ecosystems \cite{zhong2021glacier, nishimura2017environmental}. Phages play an essential role in regulating microbial system dynamics by limiting the abundance of their hosts and mediating gene transfer. For example, marine viruses can lyse 20\%-40\% of bacteria per day in marine ecosystems \cite{gregory2019marine}. In addition, by regulating the bacteria inhabiting human body sites, phages can also influence human health \cite{azimi2019phage, loc2011pros}. An important application of phage is phage therapy, which uses phages as antimicrobial agents to treat bacterial infections \cite{lee2020osong}. It has gained a resurgence of attention because of the fast rise of antibiotic-resistant bacterial infections. 

However, despite the importance of phages to both environmental and host-associated ecosystems, our knowledge about this vast, dynamic, and diverse population is very limited. Previously, the limitation is partially caused by the need of the host cell cultivation in labs. Recently, metagenomic sequencing, which allows us to obtain all genetic materials directly from a wide range of samples regardless of cultivation \cite{moon2018genomic,moon2020freshwater,moon2020viral}, has largely removed this limitation and becomes the major means for new phage discovery. According to the NCBI Reference Sequence Database (RefSeq), the number of newly released phages is doubled from 2,126 in 2019 to 4,410 in 2021. Despite the rapid growth of the phage genomes in RefSeq, the number of known phages is only the tip of the iceberg compared to those in the biome \cite{santiago2019human}. The uncharacterized phages comprise a big portion of the ``dark matter'' in metagenomic composition analysis. Due to the lack of universal marker genes, phages cannot be easily and comprehensively identified \cite{IMGVR} using conventional methods. 

There are two main challenges for phage identification in metagenomic data. First, both lytic and temperate phages can integrate the host genetic materials into their genomes, leading to local sequence similarities between the genomes of phages and bacteria \cite{edwards2016computational}. For example, $\sim$76\% phages with known hosts in the RefSeq database have detectable alignments (E-value \textless $1e\raisebox{0mm}{-}5$) with their host genomes \cite{lu2021prokaryotic}. These common regions pose challenges for distinguishing phages from their bacterial hosts. Second, although some previous works use phage structure-related genes as hallmark genes for phage identification, those genes only account for a small set of the proteins encoded by all phages. Using a small set of hallmark genes can lead to low recall of phage identification. The large gene set coded by phages is further compounded by the fact that many newly identified phages contain genes without any functional annotation. For example, \textit{Caudovirales}, the largest order of phages containing about 93\% of sequenced phages, has about 187,006 hypothetical proteins. It is not trivial to identify hallmark genes without functional annotation.

\subsection{Related work}
\label{sec:relate}
Many attempts have been made for phage identification \cite{ho2021comprehensive}. According to the algorithm design, they can be roughly divided into two groups: alignment-based \cite{VirSorter, MetaPhinder} and learning-based \cite{VirFinder, DeepVirFinder, Seeker, PPR_meta}. The alignment-based methods utilize DNA or protein sequence similarity as the main feature to distinguish phages from other sequences. For example, MetaPhinder \cite{MetaPhinder} uses BLAST hits against a phage reference database to identify phage sequences. VirSorter \cite{VirSorter} constructs a phage protein family database and applies hidden Markov model-based search to identify the protein clusters in input contigs. Then, enrichment and depletion metrics are computed to estimate the likelihood of input contigs being phages. However, the limitations of alignment-based methods are apparent. First, bacterial contigs can have multiple alignments with phage genomes, which will lead to false-positive phage predictions. Second, novel or diverged phages might not have significant alignments with the chosen phage protein families (e.g. selected hallmark genes), leading to low sensitivity for new phage identification.

To overcome the limitations of alignment-based methods, several learning-based tools have been proposed for phage identification. These learning models are mainly binary classification models with their training data containing phages and bacteria. Some learning models use extracted sequence features such as $k$-mers while others use automatically learned features in deep learning models. 
For example, VirFinder \cite{VirFinder} utilized $k$-mers to train a logistic regression model for phage detection. Virtifier \cite{miao2021virtifier} use the codon-based features to train a long short-term memory classifier for read-level phage identification. Seeker \cite{Seeker} and DeepVirFinder \cite{DeepVirFinder} encode the sequence using one-hot embedding and train a long short-term memory model and convolutional neural network, respectively. PPR-meta \cite{PPR_meta} is a three-class classification model with predictions as phages, plasmids, and choromosomes. It uses both one-hot embedding and $k$-mers to train a convolutional neural network. VirSorter2 \cite{guo2021virsorter2} employs a random forest model on sequence features, such as HMM alignment scores and GC content.

Despite the promising results, a third-party review \cite{ho2021comprehensive} shows that the precision of the these tools drops significantly on real metagenomics data. Many bacterial contigs are misclassified as phages. There are two possible reasons behind this. First, current learning models did not carefully address the challenge that phages and bacteria can share common regions. As a result, the training data did not include sufficient hard cases to train the model. For example, to construct a balanced training data, these tools often randomly select a subset of bacterial segments as negative samples. These samples may not share any local similarities with the phages and thus the trained model cannot generalize to more complicated and heterogeneous data such as real metagenomic data. Second, current models need to crop the genomes into short segments for training. The extracted features are limited to the segment and larger context information from the phage genomes cannot be effectively incorporated.

\begin{figure*}[h!]
    \centering
    \includegraphics[width=0.65\linewidth]{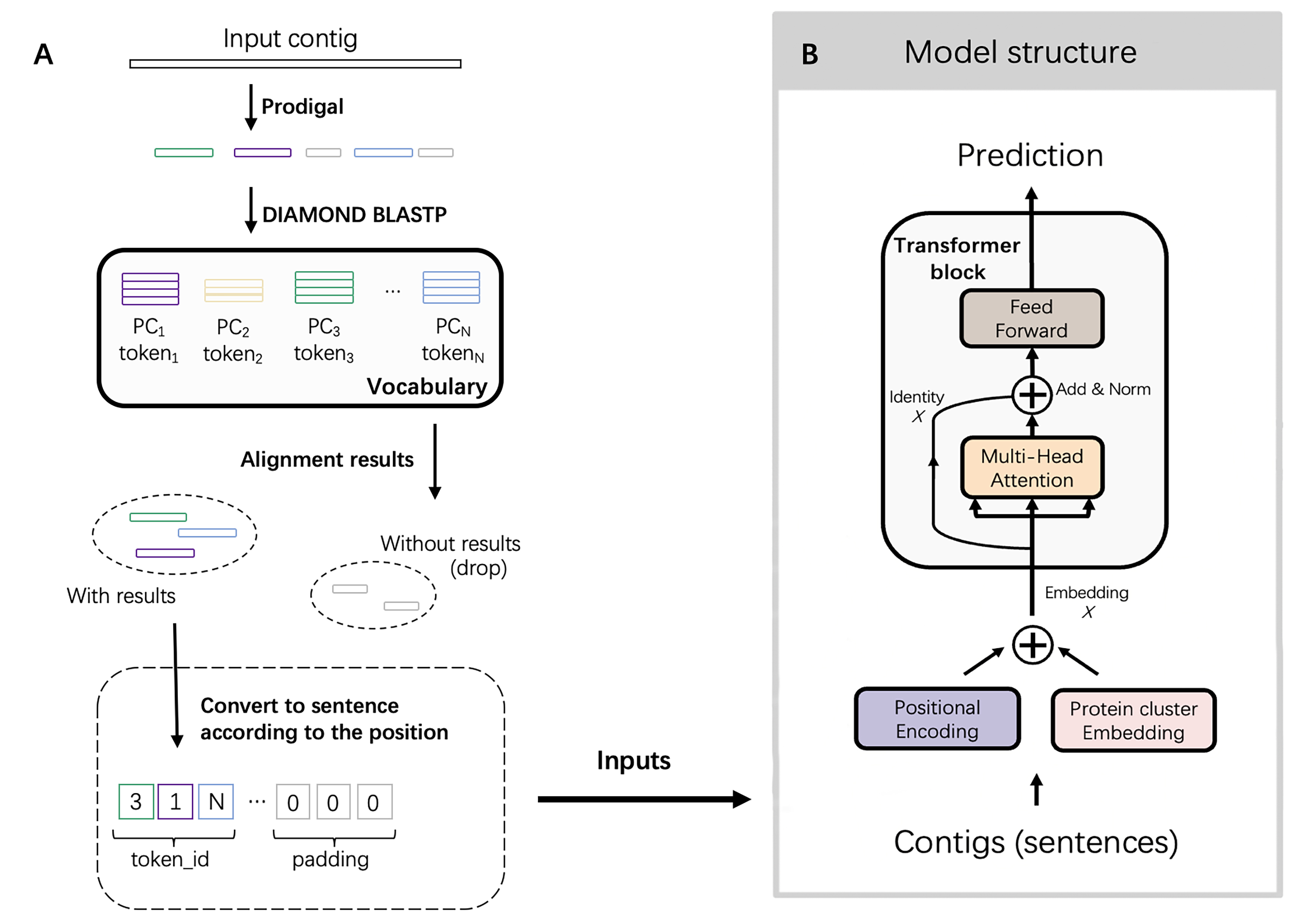}
    \caption{The pipeline of \Cherry. (A): converting inputs into protein-token sentences. During training we apply Prodigal to predict open reading frames (ORFs) from training phages and translate ORFs into proteins. Then a clustering method is applied to generate protein clusters (PCs), which are the tokens in Transformer. During the test/usage, PhaMer takes contigs as input and convert them into protein-token sentences. (B) Transformer network architecture. The converted sentences are fed into the Transformer model and a prediction is made.}
    \label{fig:pipeline}
\end{figure*}

\subsection{Overview}
In this work, we present a method, named \Cherry\, to identify phage contigs from metagenomic data.
Because previous works have shown the importance of protein composition for phage classification \cite{bolduc2017vcontact, shang2021predicting}, we employ a contextualized embedding model from natural language processing (NLP) to learn protein-associated patterns in phages. Specifically, by converting a sequence into a sentence composed of protein-based tokens, we employ the embedding model to learn both the protein composition and also their associations in phage sequences. First, we will construct the vocabulary containing protein-based tokens, which are essentially protein clusters with high similarities. Then, we apply DIAMOND BLASTP \cite{buchfink2015fast} to record the presence of tokens in training phage sequences (Fig. \ref{fig:pipeline} A). Then, the tokens and their positions will be fed into Transformer (Fig. \ref{fig:pipeline} B) for contextual-aware embedding. The embedding layer and the self-attention mechanism in Transformer enable the model to learn the importance of each protein cluster and the protein-protein associations. Although Transformer has been used for sequence embedding based on $k$-mers and motifs, we are the first in using protein clusters as tokens in Transformer for phage identification. In addition, by using the phages' host genomes as the negative samples in the training data, the model can learn from the hard cases and thus is more likely to achieve high precision in real data. Finally, \Cherry\ can directly use the whole sequences for training, avoiding the bias of segmentation. We rigorously tested \Cherry\ on multiple datasets covering different scenarios including the RefSeq dataset, short contigs, simulated metagenomic data, mock metagenomic data, and the public IMG/VR dataset. We compared \Cherry\ with four competitive learning-based tools (Seeker, DeepVirFinder, VirFinder, and PPR-meta) and one alignment-based tool (VirSorter) based on a third-party review \cite{ho2021comprehensive}. Our experimental results show that PhaMer competes favorably against the existing tools. In particular, on the mock metagenomic data, the F1-score of \Cherry\ exceeds other tools by 27\%.

\section{Methods}
Inspired by semantic analysis problems in NLP, we employ the state-of-the-art contextualized embedding model, Transformer, to automatically learn abstract patterns from the ``language'' of phages. In this language, the contigs are regarded as sentences defined on a phage-aware vocabulary. There are three major advantages behind this formulation. First, some proteins play critical roles in phages' life cycle. For example, coat proteins and receptor-binding proteins can help us distinguish phages from bacteria. These proteins can act as strong signals similar to the words describing obvious emotions in human language. Second, proteins often interact with other proteins to carry out biological functions \cite{chaban2015structural}. Similar to multiple words that can form phrases with different meanings, some protein combinations in the contigs can also provide important evidence for phage identification. Third, using protein-based tokens allows us to integrate much larger context, including the whole phage genome, into feature embedding. Unlike existing learning-based tools that often split the genomes into segments of fixed length, our model can effectively employ proteins in the whole genomes. These features prompt us to convert contigs into protein-based sentences.

In order to automatically learn the importance of proteins and their associations, we adapt the Transformer model to phage identification task. Transformer has achieved the state-of-the-art performance on a variety of NLP problems \cite{vaswani2017attention, devlin2018bert, kitaev2020reformer}. In particular, the positional encoding and self-attention mechanism enable the model to learn both the importance of each word and the relationships between words.

In the following sections, we will first introduce how we construct the protein-cluster tokens and encode the sequences into sentences. Then, we will describe the Transformer model optimized for phage identification.

\subsection{Constructing the protein-cluster tokens}
\label{sec:seq2sen}
Each token in our model is derived from a protein cluster, which contains homologous protein sequences from sequenced phages.

\paragraph{Generating protein clusters}
Our protein clusters are constructed on the training data. Specifically, they are extracted from 2,126 phage genomes released before Dec. 2018 from the RefSeq database, which constitute our training phages. More recently sequenced phages are used as test samples. Constructing the protein vocabulary using only the training sequences allows us to rigorously test our method in scenarios where newly sequenced phages harbor novel proteins outside the vocabulary. Although there are available gene annotations and their corresponding proteins for the reference genomes, we did not use the annotation. Instead, in order to be consistent with the gene prediction process of the test sequences, we apply gene finding and protein translation for the downloaded DNA genomes. According to a recent review of gene finding in viruses, Prodigal outperforms other annotation tools, especially for phages. Thus, we use Prodigal to predict ORF on both training and test genomes under its default parameters. Second, we will run all-against-all DIAMOND BLASTP \cite{buchfink2015fast} on the predicted proteins. Protein pairs with alignment E-value $\leq$ $1e\raisebox{0mm}{-}3$ are used to create a protein similarity network, where the nodes represent proteins and the edges represent the recorded alignments. The edge weight encodes the corresponding alignment's E-value. Then, Markov clustering algorithm (MCL) \cite{enright2002efficient} is employed to group similar proteins into the clusters using default parameters. All the clusters that contain fewer than two proteins are removed and finally we have 45,577 protein clusters. The size distribution of the protein clusters can be found in FigS. 1 in the [Supplementary file 1].

\paragraph{Encoding a contig into a protein token sentence}
With the generated protein clusters as the tokens in our vocabulary, we will use them to convert contigs into sentences. As shown in Fig. \ref{fig:pipeline} A, we will employ Prodigal for gene finding and translation. Then, we will identify the matched protein clusters for the translated proteins by conducting similarity search. Specifically, DIAMOND BLASP is applied to compare each translation against all the proteins in the clusters. We identify the reference protein incurring the smallest E-value and assign the query with this reference protein's cluster. We will record both the ID of the token (protein cluster) and the position of the protein in the query sequence. If an input sequence has no alignment with any token, we will not keep it for downstream analysis. Thus, if a new phage does not contain any of the tokens in our established vocabulary, it will be missed by our model and will be recorded as a false negative. In our experiments, we found that this type of phages are very rare. Most of them contain some tokens.

Because the lengths of the contigs can vary a lot, the converted sentences also contain different number of protein-cluster tokens. We follow the original paper of Transformer \cite{vaswani2017attention} and set the maximum length of the sentence to be 300. If the sequence contains more than 300 proteins clusters, we will only keep the first 300. For sequences containing less than 300 tokens, we will pad zeros at the end of the sentence. Finally, we will generate a 300-dimensional vector for the input sequence and each dimension encodes a token ID. For example, in Fig. \ref{fig:pipeline} A, we show a sentence containing three tokens: $token_1$ ($PC_1$), $token_3$ ($PC_3$), and $token_N$ ($PC_N$). The other positions in this sentence are padded with zeros. The maximum length of the sentence is a hyperparameter and can be set by users.

\subsection{The Transformer model}
The model's inputs are the converted sentences, represented by 300-dimensional vectors, and the output is a score representing how likely the input contig is a phage. The main purpose of Transformer is to automatically learn whether these sentences contain essential features for phage identification: the marker tokens (important proteins) and phrases (protein-protein associations). Two main components in Transformer contribute to these aims: 1) the embedding layers and 2) the self-attention mechanism.

\begin{figure}[h!]
    \centering
    \includegraphics[width=0.5\linewidth]{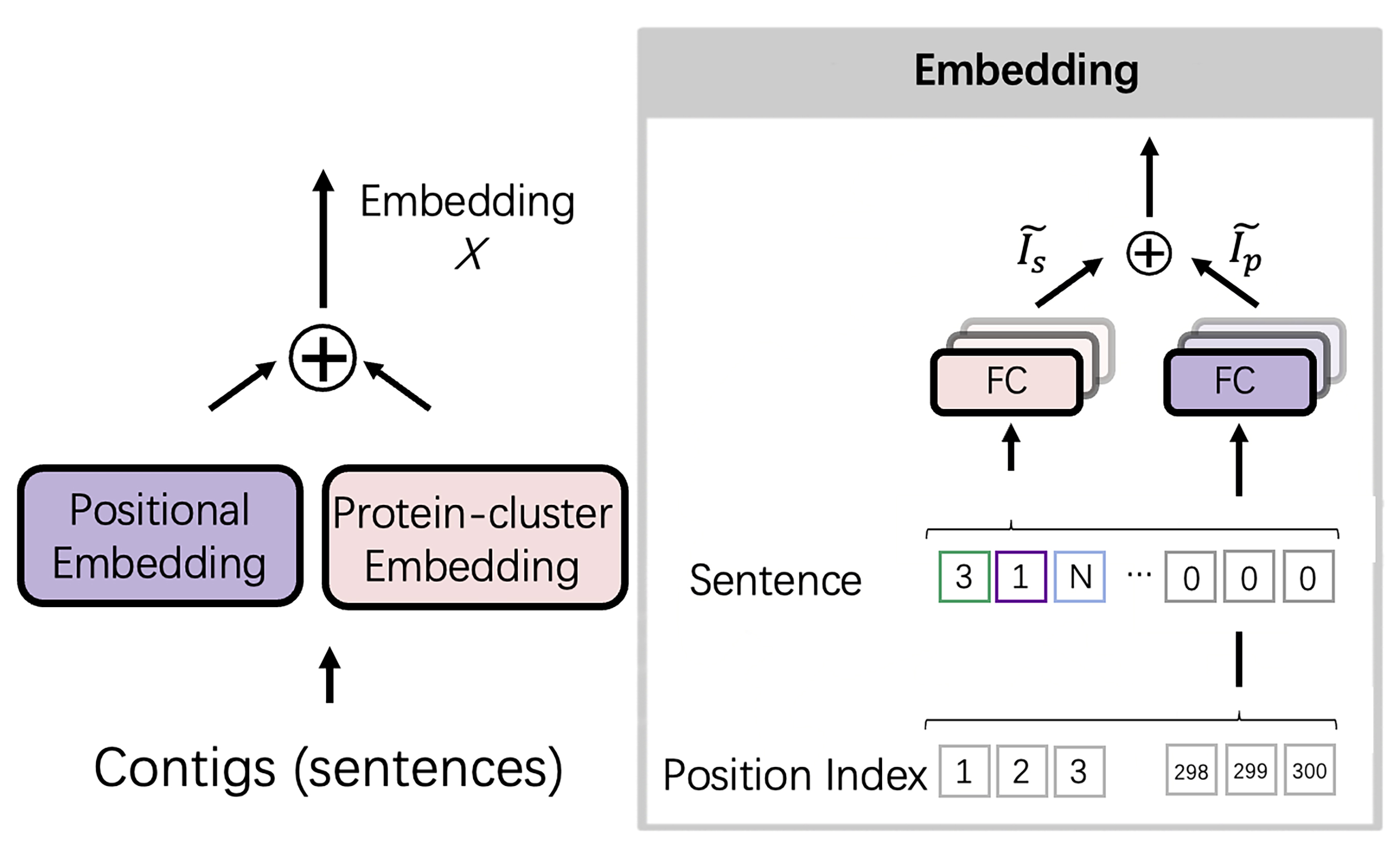}
    \caption{The embedding layer in \Cherry. There are two embedding layers in the model: protein-cluster embedding and positional embedding. The sum of these two embedding layers form the input to the Transformer block.}
    \label{fig:figure2}
\end{figure}

\paragraph{The embedding layer}
As shown in Fig \ref{fig:figure2}, before feeding the Transformer block, we embed the sentence and the position of the tokens via two embedding layers: protein-cluster embedding and positional embedding. The protein-cluster embedding layer resembles a look-up table and returns a numerical vector representing an input token. There are several ways to implement the protein-cluster embedding layer, such as the one-hot encoding used in DeepVirFinder \cite{DeepVirFinder} and Seeker \cite{Seeker}. However, because of the size of the vocabulary (45,577), using one-hot encoding can lead to very sparse vectors, which can make the model suffer from the curse of dimensionality \cite{mikolov2013distributed}. Thus, we use a fully connected layer (FC layer) to conduct linear projection for computing a low-dimensional embedding vector for each token. Because the tokens in the sentences are IDs, the mapping from the ID to a vector by the FC layer functions as a learnable dictionary (look-up table). Given an ID of a token, it will return a corresponding embedding vector of the token. 

Because Transformer contains no recurrence or convolution, it uses the positional embedding to encode the position information. The input to the positional embedding is the position index vector. The implementation is the same as the protein-cluster embedding layer with an individual learnable look-up table. The output of the embedding layer has the same dimension as the protein-cluster embedding so that the two embedded vectors can be summed. 

\begin{equation}\label{Eq1}
\left\{\begin{matrix}
\widetilde{I_s} = FC(I_s, W_{I_{s}})\\
\widetilde{I_p} = FC(I_p, W_{I_{p}})\\
X = \widetilde{I_s} + \widetilde{I_p}
\end{matrix}\right.
\end{equation}

Mathematically, the embedding layers can be presented by Eqn. \ref{Eq1}. $I_s$ is the input sentence and $I_p$ is the position index vector for the input tokens as shown in Fig \ref{fig:figure2}. $W_{I_s} \in \mathbb{R}^{N \times embed}$ and $W_{I_p} \in \mathbb{R}^{len \times embed}$ are the learnable parameters of the look-up table for protein-cluster embedding and positional embedding, respectively. $N$ is the number of protein clusters, which is 45,577 in our model, and $len$ is the maximum length of the sentence, which is 300 by default. $embed$ is a hyperparamters of the embedding dimension and it is set to 512 by default following the guideline in \cite{vaswani2017attention}. The padding tokens, which are fixed to zero, are not involved in the downstream computation. The outputs of the embedding are two matrices $\widetilde{I_s} \in \mathbb{R}^{300 \times 512}$ and $\widetilde{I_p} \in \mathbb{R}^{300 \times 512}$, where each row represents an embedded token and position vector, respectively. The final output $X \in \mathbb{R}^{300 \times 512}$ of the embedding layer is the sum of two matrices $\widetilde{I_s}$ and $\widetilde{I_p}$  Then, $X$ will be fed into the Transformer block. Ideally, these embedding layers will capture some of the semantics of the input by placing semantically similar tokens close together in the embedding space \cite{cui2018survey}. Because the value of the embedding layers in our work represents the proteins of the phages and position of proteins, the embedding could help place proteins with similar functions, such as the proteins for constructing the capsid, in proximity in the embedding space.

\begin{figure}[h!]
    \centering
    \includegraphics[width=0.6\linewidth]{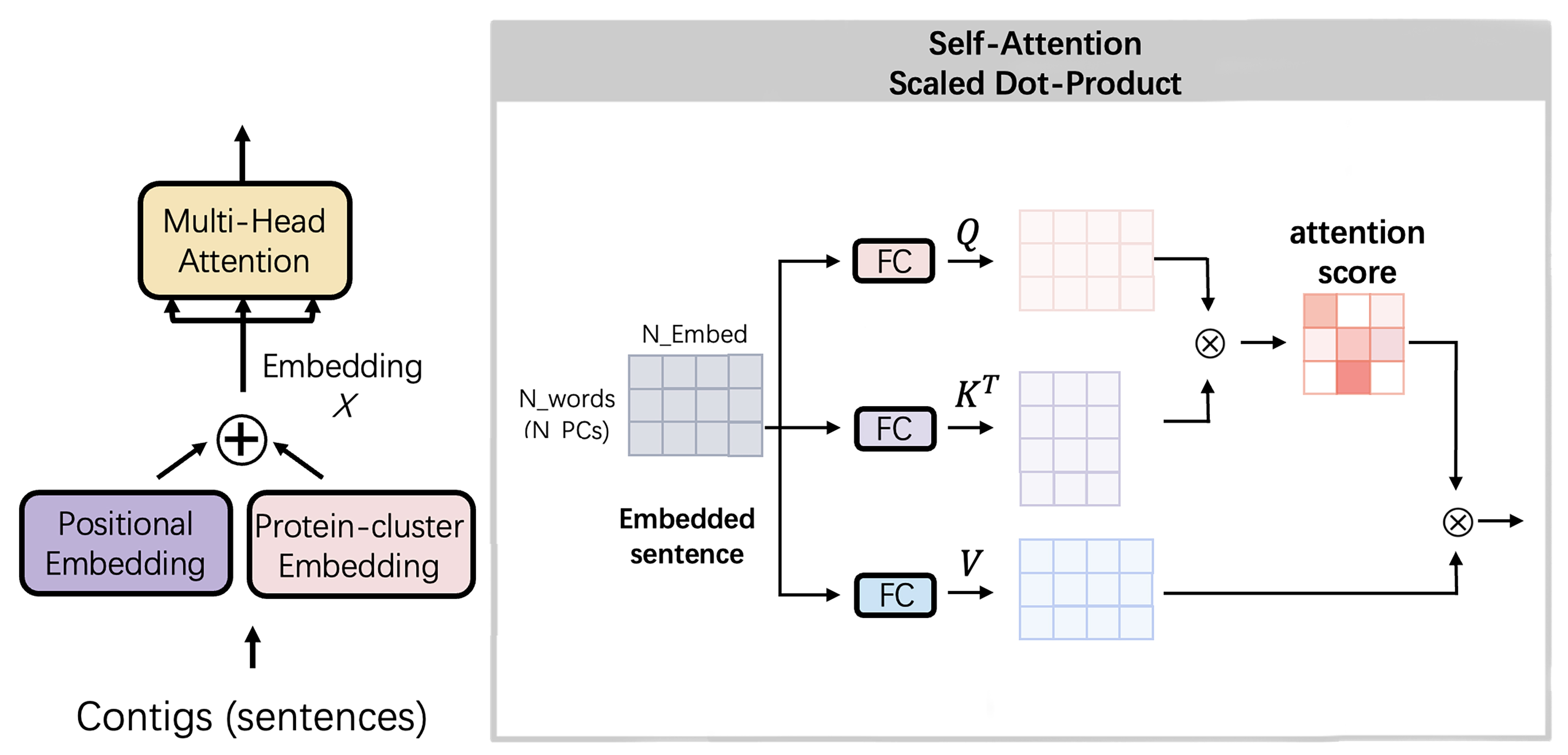}
    \caption{The self-attention mechanism in the Transformer model. The input of the self-attention is the embedded vector and the output is the weighted features with protein-protein relationships information.}
    \label{fig:figure3}
\end{figure}

\paragraph{Self-attention}
After embedding the sentences, each token is converted into a vector of size 512 and the embedded sentences will be a $\mathbb{R}^{300 \times 512}$ matrix. Then, we feed the matrix into the self-attention mechanism as shown in Fig. \ref{fig:figure3}. First, three FC layers are adopted to generate a query matrix $Q$, a key matrix $K$, and a value matrix $V$. We want to train a model to learn: given a set of proteins (query), which proteins (key) are usually co-present in phage genomes (value). Then, when making a prediction for the contigs, the model will evaluate whether the co-occurrence of some proteins shows enough evidence for phage classification.

\begin{equation}\label{Eq2}
Attention(Q, K, V) = SoftMax(\frac{QK^T}{\sqrt{d_k} } )V
\end{equation}

Fig. \ref{fig:figure3} and Eqn. \ref{Eq2} show how the self-attention mechanism works. First, the embedded matrix $X$ is projected by three FC layers into $Q$, $K$, and $V$, respectively. Second, we multiply $Q$ and the transpose of $K$ and obtain an attention score matrix of size $len$ by $len$, where $len$ is the length of the sentence. Thus, the value in the attention score matrix represents the strength of associations between two proteins. Then, the SoftMax function is employed to obtain normalized weights for each protein-cluster token and finally we multiply the weight with the value matrix $V$ to score the protein clusters in the sentences. Because the matrix multiplication between $Q$ and the transpose of $K$ might grow large in magnitude when the dimension of the embedding increases, leading to a extremely small gradients of the SoftMax function, we divided it with a scaling factor $\sqrt{d_k}$ to prevent gradient vanishing. Following the suggestion of \cite{vaswani2017attention}, we set $d_k = embed = 512$.

Because the attention matrix only contains pairwise protein cluster information, to model different combinations of pairwise relationships, we use $h$ FC layer groups for linear projections. Each group is called a head ($head_i$), and on each of these projected versions of queries $Q_i$, keys $K_i$, and values $V_i$, we can perform the self-attention mechanism in parallel. To reduce computational complexity, in each FC layer we will reduce the dimension for the projected features. The dimension of the output will be $len \times d_s$, where $d_s$ is calculated by $embed/h$. In this work we choose $h = 8$ by default. Thus, the formula of each head attention can be written in Eqn. \ref{Eq3}.

\begin{equation}\label{Eq3}
\begin{aligned}
\left\{\begin{matrix}
head_i &= Attention(Q_i, K_i, V_i)\\
Q_i &= FC(X, W^Q_i)\\
K_i &= FC(X, W^K_i)\\
V_i &= FC(X, W^V_i)\\
\end{matrix}\right.
\end{aligned}
\end{equation}

\noindent The parameters in the FC layers are projections matrices: $W^Q_i \in \mathbb{R}^{N \times d_s}$, $W^K_i \in \mathbb{R}^{N \times d_s}$, and $W^V_i \in \mathbb{R}^{N \times d_s}$. Finally, we will concatenate the output from each head and form the final output of the multi-head attention block as shown in Eqn. \ref{Eq4}, where $W^O \in \mathbb{R}^{hd_s \times embed}$.

\begin{equation}\label{Eq4}
MultiHead(Q, K, V) = FC(Concat(head_1, ..., head_h), W^O)
\end{equation}

While convolution and recurrence in CNN and RNN can record the relative positions directly, the attention mechanism is more suitable for biological data, especially for protein-cluster tokens, because the attention score can learn the remote interactions between proteins from the embedded feature. CNN and RNN can be limited by their architectures that only gives them access to local context with a limited window size or receptive field. However, the self-attention mechanism of the Transformer grants access to all positions in the embedded sentence. In addition, the positional embedding enables the model to leverage the position of each protein for prediction. Then all the information can be used in the attention mechanism simultaneously. 

\paragraph{Feed-forward networks} The output of the multi-head attention block is fed to a 2-layer neural network, which is called feed-forward networks, as shown in Fig. \ref{fig:pipeline} B. We employ a residual connection \cite{he2016deep} to the output of the multi-head, followed by layer normalization \cite{ba2016layer} to prevent overfitting. Then, we employ the $sigmoid$ function to the final output of the Transformer block to compute the probability of a contig being part of a phage.

\paragraph{Model training} During training, we first generate protein clusters and vocabulary using the phage genomes released before Dec. 2018 (i.e. our training phages). Then, we convert phage sequences into sentences using the sentence construction method. We also apply data augmentation by randomly generating short segments, ranging from 3kbp to 15kbp, to enlarge the training set. These segments are used to improve the robustness to the short contigs, which might not contain many protein clusters. We use both the segments and the complete genomes to train PhaMer to prevent the model from overfitting to the complete genomes. The training data also includes the host bacterial genomes of the training phage sequences. Because phages usually share local similarities with their hosts, we use the host genomes as the negative set to create harder negative samples for model training. Compared to using randomly selected bacterial sequences, using the hosts as the negative samples can help the model learn a more accurate classification surface and thus improves the model's generality. We download the host genomes from RefSeq database and both chromosomes and plasmids are included. The bacterial genomes go through the same sentence encoding process as the positive samples. Because some phages do not have known hosts, we sample segments from bacterial genomes to balance the training the positive and negative training data. We employ binary cross-entropy (BCE) loss and Adam optimizer with a learning rate of 0.001 to update the parameters. The model is trained on HPC with the GTX 3080 GPU unit to reduce the running time. Finally, the pre-trained model will be used to identify phages in input sequences.

\subsection{Experimental setup}
\label{sec:data}

\paragraph{Metrics}
We use the same metrics as the previous works to ensure consistency and a fair comparison: precision, recall, and F1-score. Their formulas are listed below (Eqn.\ref{m1}, Eqn. \ref{m2}, and Eqn. \ref{m3}):

\begin{equation}
    \label{m1}
    precision = \frac{ TP }{TP+FP}  
\end{equation}

\begin{equation}
    \label{m2}
    recall = \frac{ TP }{TP+FN}  
\end{equation}

\begin{equation}
    \label{m3}
    F1\raisebox{0mm}{-}score = \frac{ 2*precision*recall }{precision+recall}
\end{equation}

\noindent \textit{TP}, \textit{FN}, and \textit{FP} represent the number of corrected identified phages, missed phages, and falsely identified phages by PhaMer, respectively. If the input contigs do not contain any protein-based token, we will directly assign ``non-phage'' label to them, which become part of the \textit{FN}.

\paragraph{Dataset}
We rigorously tested \name\ on multiple datasets with increasing complexity. The detailed information is listed in Table \ref{tab:dataset}.

\begin{table*}[h]

\begin{tabular}{p{5cm}p{10.5cm}}
\hline
\textbf{Name}                 & \multicolumn{1}{c}{\textbf{Description}}                                 \\ \hline
RefSeq dataset                & We split the training and test set by time. All the phage genomes released before Dec. 2018 in RefSeq comprise the training set while the genomes released after that comprise the test set. This dataset is a widely used benchmark dataset in phage identification task. For each phage, the host information is available based on the keywords $`isolate\_host='$ or $`host='$ within each GenBank file. If no known host is available, we use this phage as a positive sample without a negative pair. Finally, 305 bacteria and 4,410 phages were downloaded. The training set contains 106 bacteria and 2,126 phages. The test set contains 194 bacteria and 2,284 phages. \\

Short contig test set         & We randomly cut the test phage genomes into segments of different lengths: 1kbp, 2kbp, 3kbp, 5kbp, 10kbp, and 15kbp. To balance the dataset, we randomly extract 10 segments from each test phage genome and 100 segments from each test bacterial genomes. Finally, we have 22,840 phage segments and 19,400 bacterial segments for each given length. Then we use these segments to evaluate the performance of phage identification on short contigs. \\

Simulated metagenomic dataset & We use a sophisticated metagenomic simulator, CAMISIM \cite{fritz2019camisim} to generate simulated data using six common bacteria living in the human gut. Instead of adding random phages to this dataset, we add simulated reads from phages that infect these bacteria to create a harder case for distinguishing phages from bacteria with shared local similarities. Then, metaSPAdes \cite{nurk2017metaspades} is applied to assemble the reads into contigs, which are fed into test phage detection tools. Finally, MetaQUAST \cite{mikheenko2016metaquast} is used to map contigs to reference phage genomes in order to assign the labels to the contigs. The experimental results can be found in section \textit{Experiments on the simulated data} in the [Supplementary file 1].              \\
Mock metagenomic dataset      & Nine shotgun metagenomic sequencing replicates of a mock community \cite{kleiner2017assessing} are retrieved from the European Nucleotide Archive (BioProject PRJEB19901). We use metaSPAdes to assemble the reads into contigs, which are used for evaluation. Similarly, The label of the contigs are determined using MetaQUAST.  \\

IMG/VR dataset                & IMG/VR v3 database \cite{IMGVR} contains 2,314,129 viral contigs assembled from different environmental samples. We recruit 354,501 contigs with known bacterial hosts. With this dataset, we will compare the recall of different tools for identifying phages from different environments. \\ \hline
\end{tabular}
\caption{The detailed information of the datasets.}
\label{tab:dataset}
\end{table*}

\section{Results}
In this section, we will show our experimental results on different datasets and compare \Cherry\ against the state-of-the-art tools. Because we want to include both alignment-based and learning-based methods in the experiments, we choose the tools that have top performance in each category based on a recent review \cite{ho2021comprehensive}. Thus, we compared one alignment-based method: VirSorter \cite{VirSorter} and four learning-based methods: Seeker \cite{Seeker}, DeepVirfinder \cite{DeepVirFinder}, VirFinder \cite{VirFinder}, and PPR-meta \cite{PPR_meta}.
Because Virtifier \cite{miao2021virtifier} is designed for short reads (length \textless 500bp) we exclude Virtifier from the comparison.

\subsection{Experiments using the RefSeq dataset}
 
\begin{figure}[h!]
    \centering
    \includegraphics[width=0.6\linewidth]{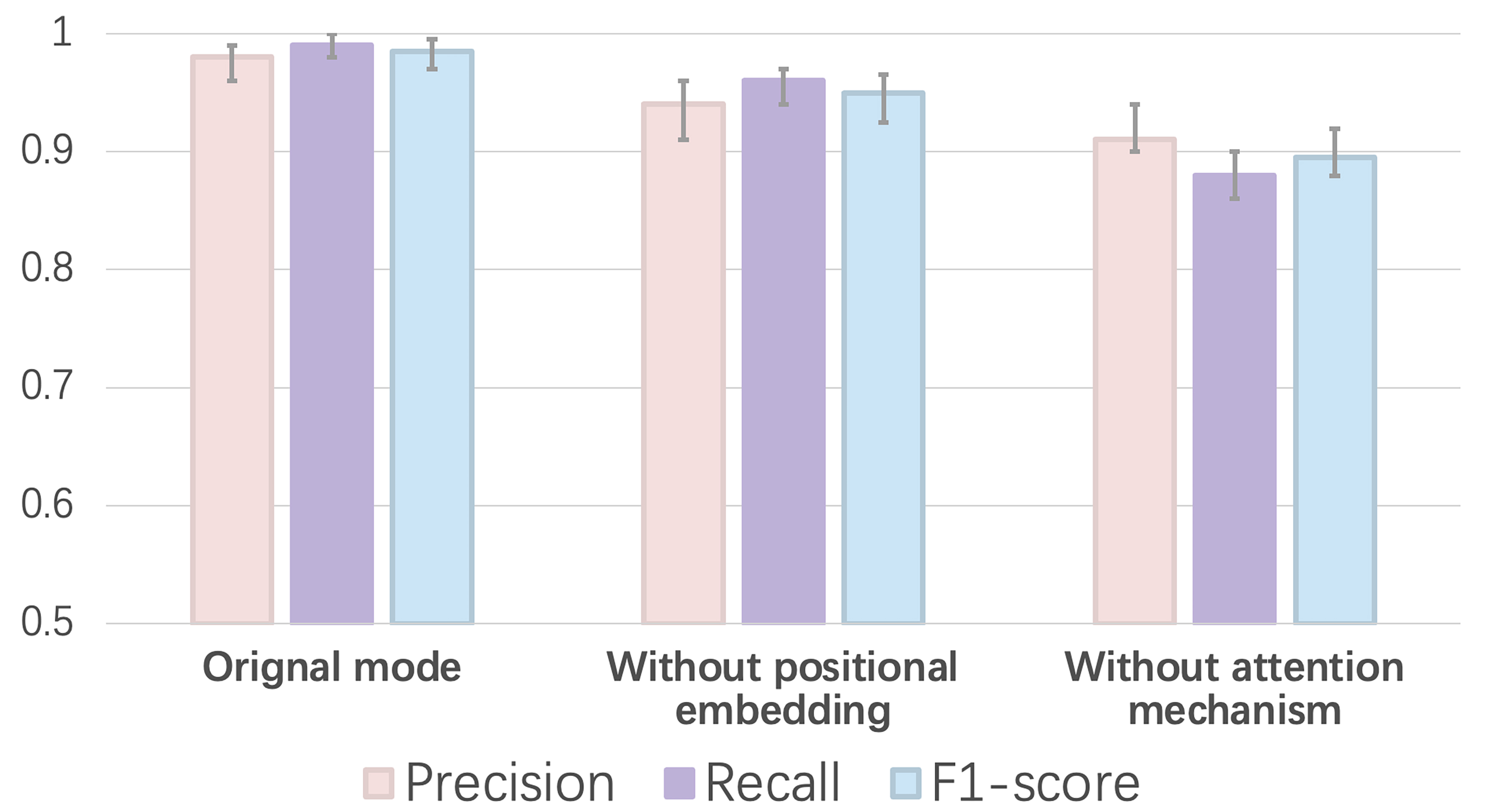}
    \caption{The ten-fold cross-validation performance on the training set. X-axis: training with different methods. Y-axis: the value of each metric (precision, recall, and F1-score).}
    \label{fig:figure4}
\end{figure}

\paragraph{Ten-fold cross-validation} We trained our model using ten-fold cross-validation. First, we split our training set into ten subsets. Second, we use nine subsets for training and the remaining one for validation. Finally, we repeated step two by iteratively choosing one subset for testing and recording the performance. The final results are shown in Fig. \ref{fig:figure4}. We also show how the positional embedding and attention mechanism affect the learning performance. \textit{Without positional embedding} means that we only use the sentence as input (shown in Fig. \ref{fig:figure2}). \textit{Without attention mechanism} means that we directly feed the embedding feature $X$ into the feed-forward network (shown in Fig. \ref{fig:figure3}). The results clearly show that both strategies improve the performance. We also visualize the attention score matrix ($QK^T \in \mathbb{R}^{len \times len}$ in Fig. \ref{fig:figure3}) to show the self-attention mechanism can learn important protein associations with biological significance. Detailed information can be found in section \textit{Visualization of the attention score} in the [Supplementary file 1]. After we conducted 10-fold cross-validation, we fixed the parameters of \Cherry\ using the model with the best performance on the validation set. The following experiments were conducted using this model, whose parameters are also available at our GitHub repository.

\begin{figure}[!h]
    \centering
    \includegraphics[width=0.6\linewidth]{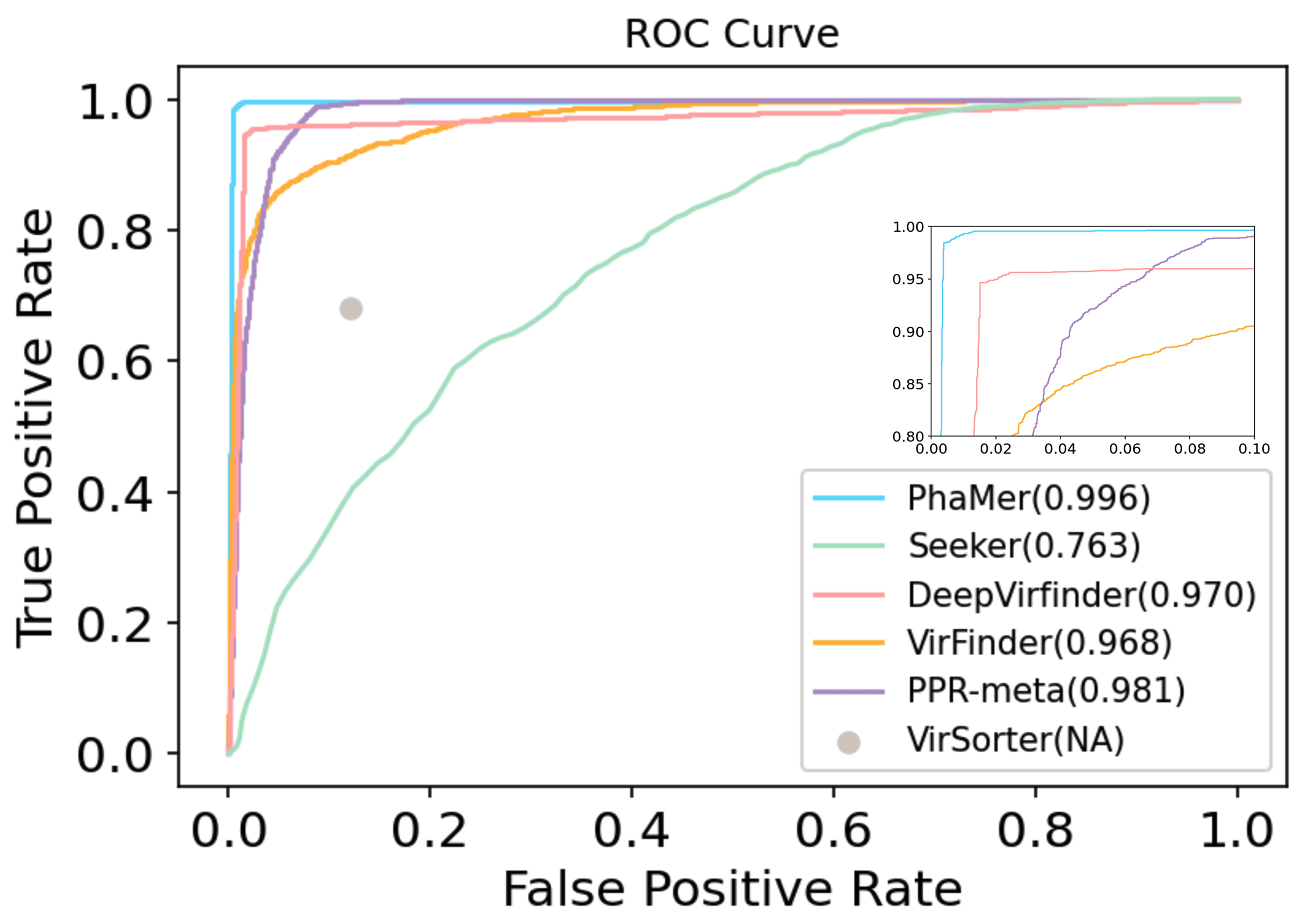}
    \caption{ROC curves on the complete test genomes. The value in the parentheses represents AUC for each tool. Because VirSorter does not provide a score associated with each prediction, we only have one data point. All later experimental results will be reported using the default cutoffs of each tool.}
    \label{fig:roc}
\end{figure}

\begin{figure}[!h]
    \centering
    \includegraphics[width=0.6\linewidth]{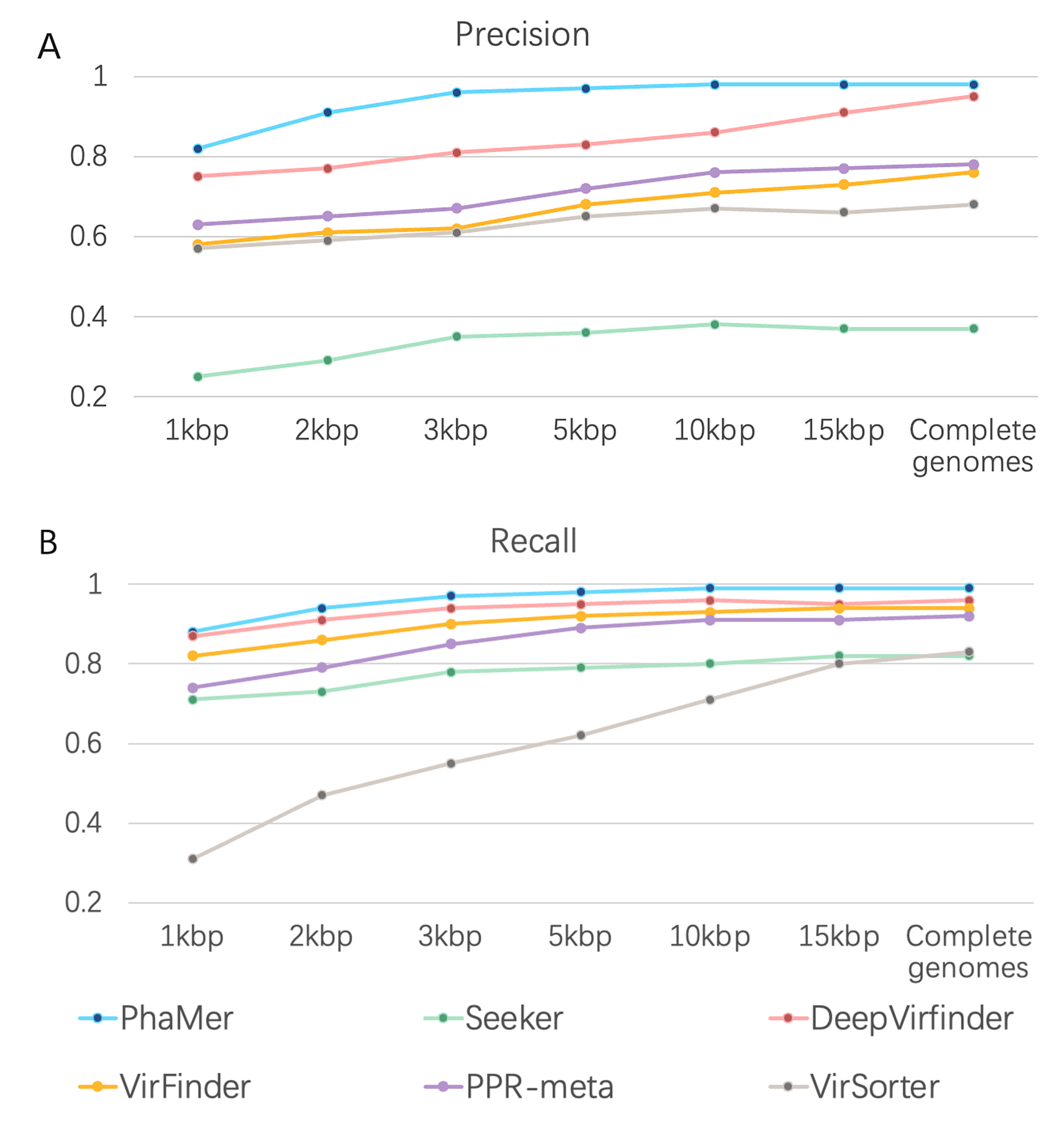}
    \caption{Precision and recall on the genomes and the simulated contigs from the test phages and bacteria. X-axis: contigs with different length. Y-axis: the precision on the test set (A) and the recall on the test set (B). For simulated contigs, there are 22,840 phage contigs and 19,400 bacterial contigs for each length range. The reported performance is averaged on three such sets of contigs for each length range. The precision and recall of all tools correspond to their default score cutoffs (0.5).}
    \label{fig:figure6}
\end{figure}

\paragraph{Performance on the test set and short contigs} We run all the state-of-the-art methods using the pre-trained model with the default parameters on the test sequences. For complete input genomes as inputs, we draw a ROC curve using the output score of each tool in Fig. \ref{fig:roc}. The area under the ROC curve reveals that PhaMer has more reliable results on the complete phages. Then, under the same score cutoff value (0.5) as all other tested tools, we recorded their precision and recall in Fig. \ref{fig:figure6}. Meanwhile, we tested the tools on the \textit{short contig test set} described in Section \textit{Experimental setup}. To reduce the bias of data generation, we repeated the short contig generation process for three times and reported the average performance in Fig. \ref{fig:figure6}.  The comparison reveals that \Cherry\ can achieve the best performance across all length ranges. With the increase of contig length, the performance of all pipelines increases. This is expected because longer sequences may contain more information for phage identification. 
As Fig. \ref{fig:figure6} shows, when the contigs are as short as 1kbp, \Cherry\ has precision around 0.8 while others have precision lower than 0.8. Thus, we do not suggest that users conduct phage identification for even shorter contigs, which can lead to unreliable results.

Because DeepVirFinder and Seeker support training a customized model, we also tried to re-train these methods with our training set. However, the recall of DeepVirFinder and Seeker will dropped to 0.47 and 0.65 on the complete genomes, respectively. This indicates the possibility that their training set might contain some of our test genomes. To keep the better results, we only reported the predictions by the pre-trained models in all the experiments.

\begin{figure}[h!]
    \centering
    \includegraphics[width=0.6\linewidth]{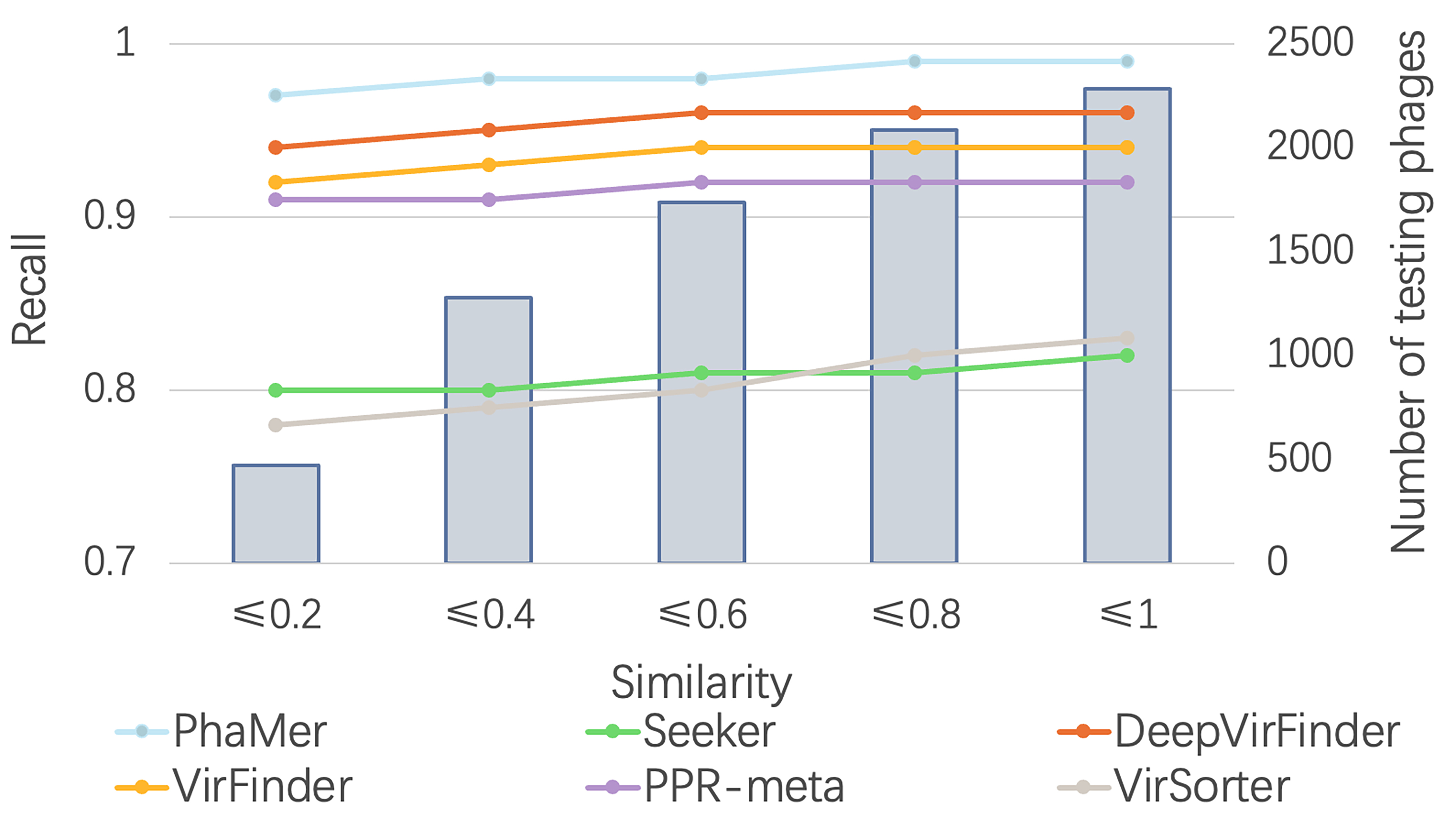}
    \caption{The impact of training-vs-test phage similarity on the recall of PhaMer. X-axis: the dashing similarity between test and training phages. The line plot: recall (left Y-axis). The bar plot: number of test phages (right Y-axis). }
    \label{fig:similarity}
\end{figure}

\paragraph{The similarity between the training and test set} The performance of phage identification can be affected by the similarity between the training and test sequences. We used Dashing \cite{baker2019dashing} to estimate the similarity between the training and test set. First, we ran all-against-all comparisons between sequences in the training and test set. Then, we recorded the largest similarity for each test phage in the test set. The mean value of the similarity is 0.41, indicating a relatively low similarity between the test and training set. To show how the similarity affects the phage identification, we divided the test set according to the dashing similarity. Because low similarity between training and test phages mainly affects the recall of phage identification, we recorded the recall in Fig. \ref{fig:similarity}. X-axis stands for the maximum similarity between genomes in the training and test sets. For example, X-axis value 0.2 indicates that all the test phages have similarity $\leq$ 0.2 against the training phages. Fig. \ref{fig:similarity} shows that with the increase of the similarity, the recall of all methods increases and \Cherry\ outperforms other tools on a wide range of similarities.

\subsection{Experiments on the mock metagenomic data}
\begin{figure*}[h!]
    \centering
    \includegraphics[width=0.7\linewidth]{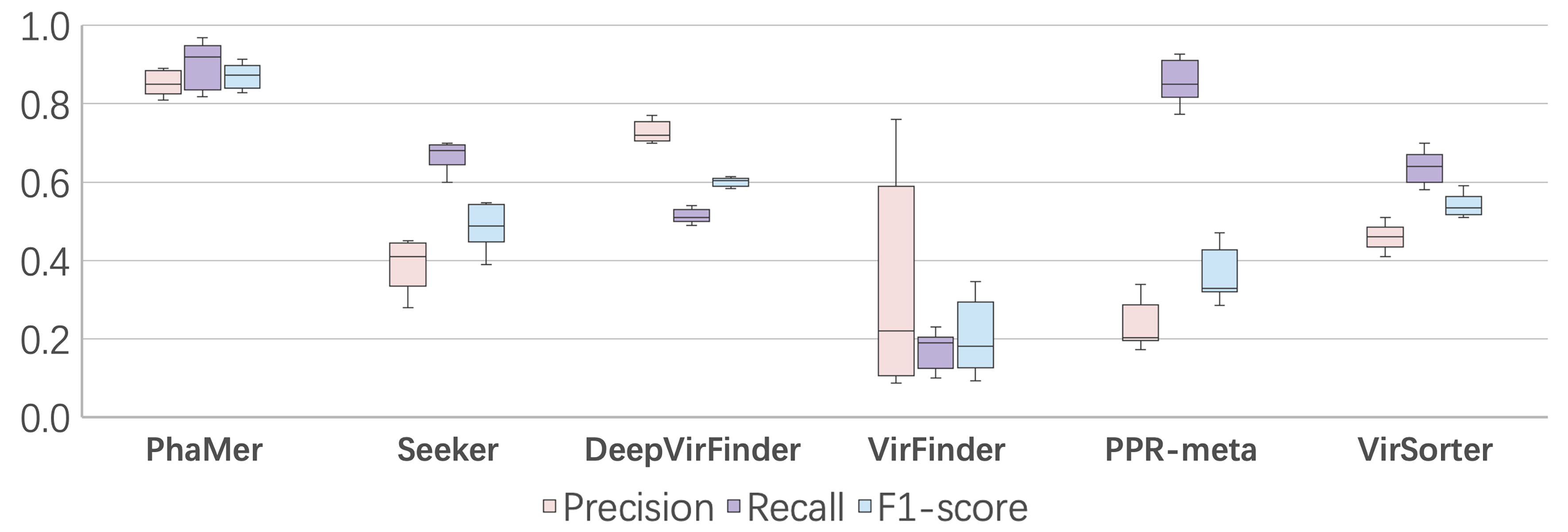}
    \caption{Results on the mock metagenomic data. X-axis: the names of the compared methods. Y-axis: the score of the metrics. Results on the mock metagenomic data. PhaMer outperforms the other tools regarding precision, recall and F1-score. PPR-meta achieves the second-best recall but shows low precision.}
    \label{fig:figure8}
\end{figure*}

After validating \Cherry\ on the RefSeq database and the simulated datasets, we compare all the methods on real shotgun-sequenced metagenomic datasets that are used for testing phage identification by the review \cite{ho2021comprehensive}. The sequencing data are from a mock community \cite{kleiner2017assessing}, which contains 32 species or strains, including 5 phages. There are 9 sequenced datasets with different properties of cell number abundance and protein biomass level from this mock community. These datasets are publicly available at European Nucleotide Archive (BioProject PRJEB19901). Following the guidelines of \cite{ho2021comprehensive}, we used the FASTQC \cite{andrews2017fastqc} to control the quality of the data and removed over-represented reads with Cutadapt \cite{martin2011cutadapt}. The cleaned paired-end reads were fed into metaSPAdes \cite{nurk2017metaspades} and the output contigs were labelled by MetaQUAST \cite{mikheenko2016metaquast}. Only the contigs with length \textgreater 3kbp will be used for comparison and the prediction results of all methods are shown in Fig. \ref{fig:figure8}.

In general, the F1-scores of other tools were considerably lower on this dataset than on the RefSeq benchmark dataset, with an average F1-scores drop by $\sim$30\%. A closer look shows that they commonly misclassified the bacterial contigs as phages in this metagenomic data. The precision of \Cherry\ is still much better than the state-of-the-art methods. Because \Cherry\ learns not only the importance of proteins but also the associations between proteins from phage sequences, it is able to make a fine distinction between bacteria and phages. In addition, we use the hard cases (the host genomes) for training, enabling the model to generalize to real metagenomic data. 

\subsection{Experiments on the IMG/VR data}

Recently, IMG/VR published the largest public virus genome database IMG/VR v3 \cite{IMGVR}. The viruses in this database are quality checked, taxonomically classified, and annotated. This dataset provides a good test set to evaluate the recall of phage identification tools. Following the previous work \cite{Seeker}, we downloaded 2,314,129 viral contigs assembled from different environmental samples and recruited 354,699 viral contigs with known bacterial hosts. All contigs shorter than 3kbp were removed, resulting in 354,501 phage contigs. Such a broad coverage of phages from different environmental niches allow us to test the reliability of the phage identification tools. Because only phages were tested in the experiments, we reported the recall of different tools.

\begin{figure}[h!]
    \centering
    \includegraphics[width=0.7\linewidth]{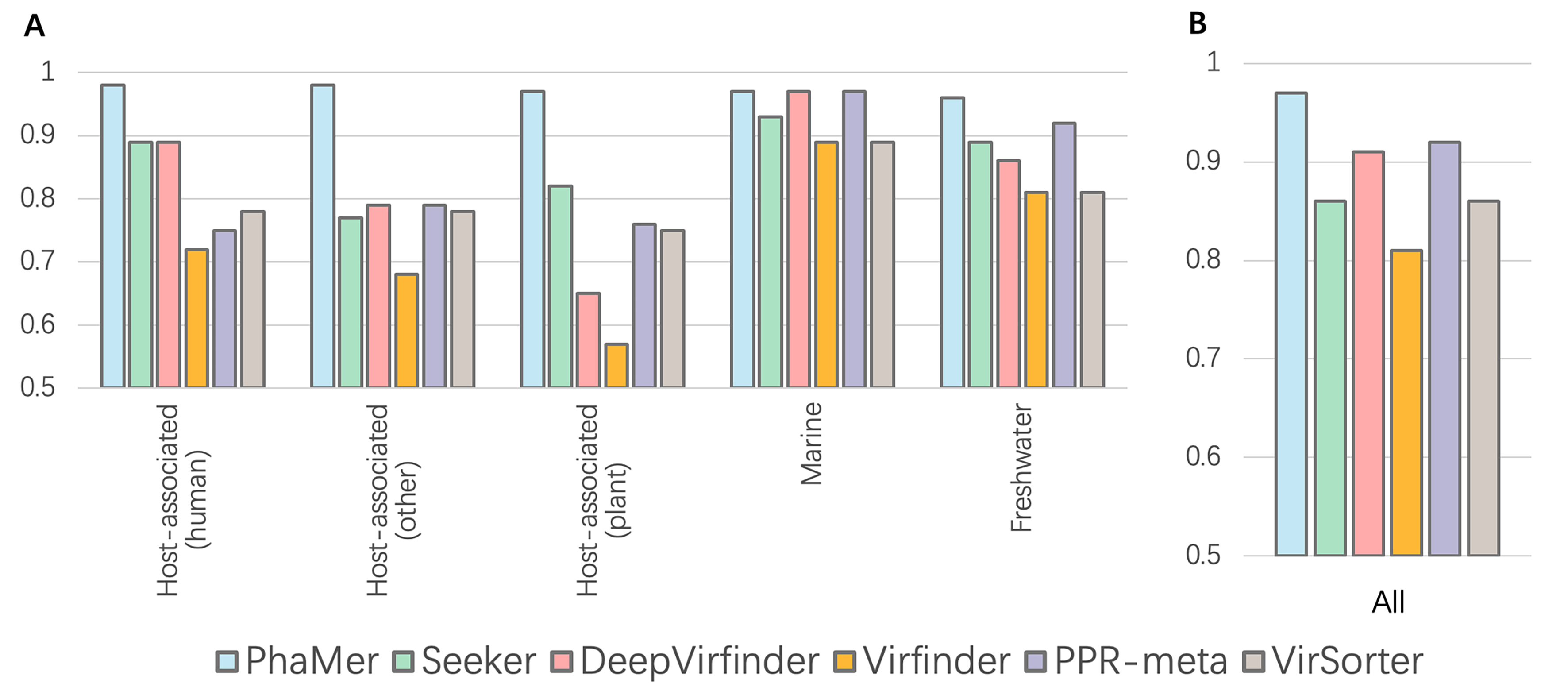}
    \caption{Results on the IMG/VR v3 database. Y-axis: the recall of \Cherry\ and the state-of-the-art tools for identifying phages in the IMG/VR database. A: Recall of different tools on phages living in different environments; B: the overall recall on the whole database.}
    \label{fig:figure9}
\end{figure}

Fig. \ref{fig:figure9} shows the recall of all six methods on the IMG/VR dataset. We split the dataset according to the living environments of the phages and show the performance of all tools on five of them with largest data samples (Fig. \ref{fig:figure9} A): \textit{human}, \textit{plant}, \textit{marine}, \textit{freshwater}, and \textit{other}. The results reveal that \Cherry\ outperforms all other methods on these five domains. In particular, \name\ significantly improved the recall of phage identification in plant-associated samples. Fig. \ref{fig:figure9} B shows the summarized recall on all the datasets with PhaMer having a recall of 5\% higher than the second best tool PPR-meta. 

\section{Running time comparison}

The most resource-demanding components in \Cherry\ are the translation (Prodigal) and sequence alignment (DIAMOND BLASTP). We used these two steps to convert input sequences into protein-based sentences. Both the protein cluster vocabulary and position information are utilized as input features of the Transformer model. Table \ref{tab:time} shows the average elapsed time of classifying the test set (2,284 genomes) for each tool. \Cherry\ is not the fastest program, and $\sim$90\% running time is used to run Prodigal and DIAMOND BLASTP.

\begin{table}[h!]
\centering
\begin{tabular}{lllllll} \hline
Program                       & \Cherry\ &  VirFinder & DeepVirFinder & Seeker & PPR-meta & VirSorter \\ \hline
Elapsed time(min) & 67     & 31           & 23   & 58     &     46 &  214\\ \hline
\end{tabular}
\caption{The average elapsed time to make predictions for the RefSeq test genomes. All the methods are run on Intel\textsuperscript{\textregistered} Xeon\textsuperscript{\textregistered} Gold 6258R CPU with 8 cores.}
\label{tab:time}
\end{table}

\section{Discussion}
As shown in the experiments, existing approaches, such as Seeker, VirFinder, DeepVirFinder, PPR-meta, and VirSorter, failed to achieve high precision or recall in phage identification, especially on metagenomic data. In this work, we demonstrate that \Cherry\ can render better performance for novel phage identification. The major improvement of our method stems from the adoption of the language model Transformer for bidirectional contextual protein embedding and our careful construction of negative samples. By constructing protein-based vocabulary, we can incorporate the similarity between phages. By using the positional embedding and the self-attention mechanism in Transformer, we can learn the importance of proteins and also their associations. In addition, we carefully construct our negative training samples using phages' host bacterial genomes. Because of the local similarities between phages and their hosts, this negative training set is more difficult to classify and tend to have direct bearing on learning the optimal decision surface. The benchmark experimental results on complete genomes, short contigs, simple and complicated metagenomic data show that our model outperforms others in different scenarios. Importantly, its performance is more robust than others on short contigs. And it increases the F1-score by 27\% on the mock metagenomic data.

Considering that we first constructed protein clusters from phage proteins, it is a fair question to ask whether we can achieve an optimal classification by controlling the protein matching criteria (E-value). When constructing our negative set (bacterial host sequences), we found that over 80\% of bacterial genomes have multiple alignments with the protein clusters in our vocabulary. Thus, there is a tradeoff between the recall and precision in terms of the E-value cutoff. Because phages are highly diverse and some protein clusters can have remote homologs in new phages, using a very stringent E-value cutoff can miss new pages. We found that if we use a very strict E-value cutoff (e.g., E-value close to 0), most of these bacterial genomes can be rejected. However, the recall of identifying phages will drop to 0.3. Thus, it is hard to strike an optimal balance between sensitivity and precision by adjusting E-value cutoffs.

Although \Cherry\ has greatly improved phage contig identification, we have several aims to optimize \Cherry\ in our future work. One possible extension is to incorporate  prophage detection in PhaMer. There are a number of prophage annotation tools, such as Prophage Hunter and Phage\_Finder. In our software, we supply a metric named \textit{proportion} to record the ratio of matched tokens in our vocabulary to the predicted proteins by Prodigal for each input contig. Intuitively, for each input that is predicted as ``phage'' by PhaMer, we can further check the value of \textit{proportion}. If this value is very small, the input contig is likely to be a prophage rather than a phage. Users can filter the contigs according to their needs conveniently. In the future, we will incorporate prophage detection tools in our pipeline.

Like a majority of phage identification tools, PhaMer is a binary classification model with the goal of distinguishing phages from bacterial contigs, which can originate from either chromosomes or plasmids.
In our current design and test, we have plasmid sequences in both the training and test sequences. For example, some contigs assembled from the mock metagenomic data are from plasmids. Because the plasmid-originated sequences are treated as ``negative'' samples as the bacterial chromosomes, PhaMer does not distinguish them from chromosomes. There are other tools available to distinguish plasmids from the host genomes. Users can run those tools for downstream analysis.

%
%


\section*{Funding}
City University of Hong Kong (Project 9678241), HKIDS (9360163), and the Hong Kong Innovation and Technology Commission (InnoHK Project CIMDA).

\bibliographystyle{unsrt}  
\bibliography{references}  

\end{document}